\documentclass[aps,prx,reprint,twocolumn,superscriptaddress,english]{revtex4}
\usepackage{amssymb}
\usepackage{graphicx}
\usepackage{amsmath}
\usepackage{epsfig}

\usepackage[latin9]{inputenc}
\usepackage{array}
\usepackage{multirow}
\usepackage[monochrome]{color}

\usepackage{esint}
\usepackage{bm}
\usepackage{bbm}
\usepackage{hyperref}
\usepackage{babel}
\usepackage{epstopdf}
\usepackage{mathtools}
\usepackage{soul}
\usepackage[dvipsnames]{xcolor}
\usepackage{tikz}
\begin{document}

\setcounter{equation}{0} \setcounter{figure}{0}
\setcounter{table}{0} \setcounter{page}{1} \makeatletter

\global\long\def\id{\mathbbm{1}}
\global\long\def\ui{\mathbbm{i}}
\global\long\def\ud{\mathrm{d}}

\title{Evidence for Bosonization in a three-dimensional gas of SU($N$) fermions}

\author{Bo Song}
\thanks{These authors contributed equally to this work.}
\affiliation{Department of Physics, The Hong Kong University of Science and Technology,\\ Clear Water Bay, Kowloon, Hong Kong, China}

\author{Yangqian Yan}
\thanks{These authors contributed equally to this work.}
\affiliation{Department of Physics and Astronomy, Purdue University, West Lafayette, Indiana 47907, USA}

\author{Chengdong He}
\affiliation{Department of Physics, The Hong Kong University of Science and Technology,\\ Clear Water Bay, Kowloon, Hong Kong, China}

\author{Zejian Ren}
\affiliation{Department of Physics, The Hong Kong University of Science and Technology,\\ Clear Water Bay, Kowloon, Hong Kong, China}

\author{Qi Zhou}
\email{zhou753@purdue.edu}
\affiliation{Department of Physics and Astronomy, Purdue University, West Lafayette, Indiana 47907, USA}
\affiliation{ Purdue Quantum Science and Engineering Institute, Purdue University,
1205 W State St, West Lafayette, West Lafayette, IN 47907, USA}

\author{Gyu-Boong Jo}
\email{gbjo@ust.hk}
\affiliation{Department of Physics, The Hong Kong University of Science and Technology,\\ Clear Water Bay, Kowloon, Hong Kong, China}

\date{\today}
\begin{abstract}
Blurring the boundary between bosons and fermions lies at the heart of a wide range of intriguing quantum phenomena in multiple disciplines, ranging from condensed matter physics  and atomic, molecular and optical physics to high energy physics. One such example is a multi-component Fermi gas with SU($N$) symmetry that is expected to behave like spinless bosons in the large $N$ limit, where the large number of internal states weakens constraints from the Pauli exclusion principle. However, bosonization in SU($N$) fermions has never been established in high dimensions where exact solutions are absent. Here, we report direct evidence for bosonization in a SU($N$) fermionic ytterbium gas with tunable $N$ in three dimensions (3D). We measure contacts, the central quantity controlling dilute quantum gases, from the momentum distribution, and find that the contact per spin approaches a constant with a 1/$N$ scaling in the low fugacity regime consistent with our theoretical prediction. This scaling signifies the vanishing role of the fermionic statistics in thermodynamics, and allows us to verify bosonization through measuring a single physical quantity. Our work delivers a highly controllable quantum simulator to exchange the bosonic and fermionic statistics through tuning the internal degrees of freedom in any generic dimensions. It also suggests a new route towards exploring multi-component quantum systems and their underlying symmetries with contacts. 
\end{abstract}

\maketitle
% \tableofcontents
\newpage
\section{Introduction}
Bosons and fermions exhibit intrinsically different properties because of the distinct underlying statistics. Strikingly, the boundary between bosons and fermions could become blurred under a variety of  scenarios in condensed matter and high energy physics~\cite{Weinberg:2000to,Giamarchi:2003uc,2018sci,Cherman:2012dz}, ranging from the supersymmetry exchanging bosons and fermions~\cite{2018sci} to fermionization of strongly interacting bosons in 1D~\cite{Weinberg:2000to,Giamarchi:2003uc}. In the latter case, hardcore bosons and noninteracting fermions share identical thermodynamical properties despite the fact that their correlation functions are different~\cite{paredes2004,kinoshita2004}. %%Multi-component fermions with the SU($N$) symmetry provide physicists with another rich playground to explore the interplay between bosons and fermions~\cite{Yang:2011fp}.
  Another interesting route is to increase the number of spin component $N$ in SU($N$) fermions leading to bosonization~\cite{Yang:2011fp}. Theoretically, such bosonization of SU($N$) fermions has been extensively studied in 1D~\cite{Yang:2011fp,guan:2012,liu2014,Jiang2016,Jen2018,1DRMP,Decamp2016,laird2017}.  In this particular reduced dimension, exact solutions exist and allow one to confirm bosonization in the large $N$ limit~\cite{Yang:2011fp,guan:2012}. Experimentally, this phenomenon has also been explored in 1D showing that the breathing mode of SU($N$) fermions approaches that of bosons with increasing $N$~\cite{Pagano:2014hy}. 
  
In spite of the aforementioned serious efforts of studying bosonization of  SU($N$) fermions, some fundamental questions about bosonization of SU($N$) fermions remain unanswered so far. First, does bosonization of SU($N$) fermions occur in high dimensions?  Since exact solutions generically do not exist beyond 1D, it is challenging to rigorously prove the bosonization in high dimensions. In addition, the breathing mode alone cannot tell whether other thermodynamic quantities approach those of bosons.  In practice, it is difficult to measure all thermodynamic quantities. Therefore, is it possible to use a single quantity to establish bosonization?

 In this work, we explore bosonization of a 3D SU($N$) Fermi gas by measuring its central quantity, the so-called contact, $\mathcal{C}$~\cite{Tan:2008ey, Tan:2008eg,Tan:2008ch} and have answered both questions. Through  celebrated universal relations, contacts govern  other physical observables, such as the momentum distribution, the energy, the pressure, and a variety of spectroscopies~\cite{Patridge2005,Werner2009,Stewart:2010fy,Kuhnle2011,Fletcher2017,Laurent2017}.  Therefore, the dependence of contacts on $N$ directly provides us with the evidence of bosonization without resorting measuring other thermodynamic quantities. We choose $^{173}$Yb atoms as our sample,
%%In this work, we address contacts in degenerate $^{173}$Yb atoms, which provides a direct access to many-body properties of the SU($N$) system.}
%%Such alkaline-earth-like fermions have unique features distinguishing themselves from their alkaline counterparts. First,
in which the number of %the internal degree of freedom 
internal states accessible in experiments is highly tunable, ranging from one to six. Due to the strong decoupling between electronic and nuclear spins, interactions between nuclear spins are isotropic, providing the many-body system with a SU($N$) symmetry and consequently, a wide range of exotic phenomena~\cite{Wu:2003es,Hermele:2009ev,Gorshkov:2010hw,Cazalilla:2014kq}. %%%These two characteristics of $^{173}$Yb atoms allow us to explore how contacts determine the thermodynamical quantities, such as the momentum distribution, in large spin systems, and meanwhile unfold the underlying symmetry of the many-body system.
%tunable spin $N$ and explore the scaling law of contact with $N$. Our observation should provide an important insight into the concept of contact in a large spin Fermi system.

%A high spin Fermi gas is exemplified in alkaline-earth-like fermions (e.g. ytterbium or strontium) that reveal spin-independent scattering property due to the strong decoupling between electronic and nuclear spins. So far, however, the emergence of SU($N$) symmetric interactions have been indirectly investigated in the presence of  strong correlation imposed by deep optical lattices~\cite{Taie:2012tb,Pagano:2014hy,Hofrichter:2016iq} or Feshbach resonances. Alternatively, the SU($N$) effect has been observed in a spectroscopy~\cite{Zhang:2014el} and in collective excitations~\cite{He:2019ti}.

%In this work, we report an experimental measurement of the $s$-wave contact parameter by recording the high-momentum tail of SU($N$) fermions of degenerate $^{173}$Yb atoms. 

%%%{\color{red}The SU($N$) symmetry offers multi-component fermions a unique property of bosonization in the large $N$ limit~\cite{Yang:2011fp} as the Pauli blocking becomes less effective for large spin systems. So far, however, the bosonization has been observed only in a 1D SU($N$) system where the concept of quantum statistics is not well-defined in the strong coupling regime. Here, we experimentally show that the two-body contact reveals bosonization in a 3D SU($N$) system in good agreement with theoretical results.} 

Whereas the SU($N$) symmetry has been explored in optical lattices~\cite{Taie:2012tb,Pagano:2014hy,Hofrichter:2016iq,Ozawa:2018ge}, a spectroscopy~\cite{Zhang:2014el,Cappellini2014,Scazza2015}, and collective excitations~\cite{Pagano:2014hy,He:2019ti}, it is still challenging to measure the rather small contact due to the weak interactions between $^{173}$Yb atoms. To overcome this obstacle, we develop a new protocol to extract the contact from the column integrated momentum distribution without using the inverse-Abel transform, which allows a high signal-to-noise ratio (SNR). We %observe that the contact increases as $(T/T_F)^{-3/2}$, 
{\color{blue} measure the temperature dependence of the contact} 
when the temperature $T/T_F$ decreases from 1.0 to 0.55, {\color{blue} and compare experimental results with theoretical calculations based on the virial expansion. Whereas the second order virial expansion shows a scaling of $(T/T_F)^{-3/2}$, high order virial coefficients lead to corrections from other powers of $T/T_F$. } %In particular, we find a power-law dependence of the contact on the temperature, i.e.,  $\mathcal{C}\propto (T/T_F)^{-3/2}$. 
When $N$ is fixed, no change in the measured contact is observed for different spin constituents, confirming the isotropic interaction.  We emphasize that the underlying mechanism for SU($N$) fermions, the large internal degree of freedom weakening the Pauli exclusion principle, is the same for any temperatures and any interaction strengths.  In different parameter regimes, the quantitative difference is how fast physical observables approach those of bosons. We thus focus on the temperature regime readily achievable in current experiments to deliver evidence for bosonization in three dimensions for the first time in laboratories, though the multi-component nature of SU($N$) fermions may allow a more efficient cooling down to even lower temperatures~\cite{Sonderhouse: 2020}.  Since only a finite $N$ is accessible in realistic experiments, it is critical to work out and experimentally verify how physical observables scale with $N$ so as to access an unambiguous proof of bosonization in the large $N$ limit. To this end, 
we further change the number of nuclear spin component $N$ and keep the number of atoms per component constant at the same temperature and trap geometry. We find a linear dependence of the contact with $N$.  Consequently, the contact per spin approaches a constant with a scaling law of $1-1/N$.  %%which is a direct  manifestation of the SU($N$) symmetry of our system and %%also unfolds its intrinsic property of bosonization in the large $N$ limit. 

  %without changing the number of atoms per component. 

\begin{figure}[tbp] %[tbp]
\includegraphics[width=0.9\linewidth]{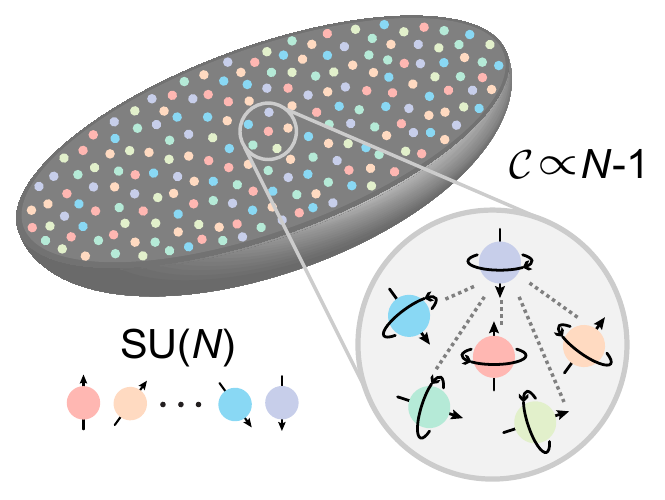}
\centering
\caption{\textbf{Illustration of %Tan's 
 $s$-wave contacts in SU($N$) fermions with tunable spin}. Arrows with different colors and orientations denote the different nuclear spin states as large as $N$=6. Dashed lines represent pairs formed by two particles with different spins. Each pair contributes equally to the contact, which leads to $\mathcal{C}\propto N-1$.}

%%Tan's contact characterizes the short-range interaction and it follows the scaling law that contact per particle $\mathcal{C}$ in a spin balanced gas of $N$ spin states, is proportional to $(N-1)$.}
	\label{fig1_scheme}
\end{figure}

\begin{figure*}[!htb]
	\includegraphics[width=0.75\linewidth]{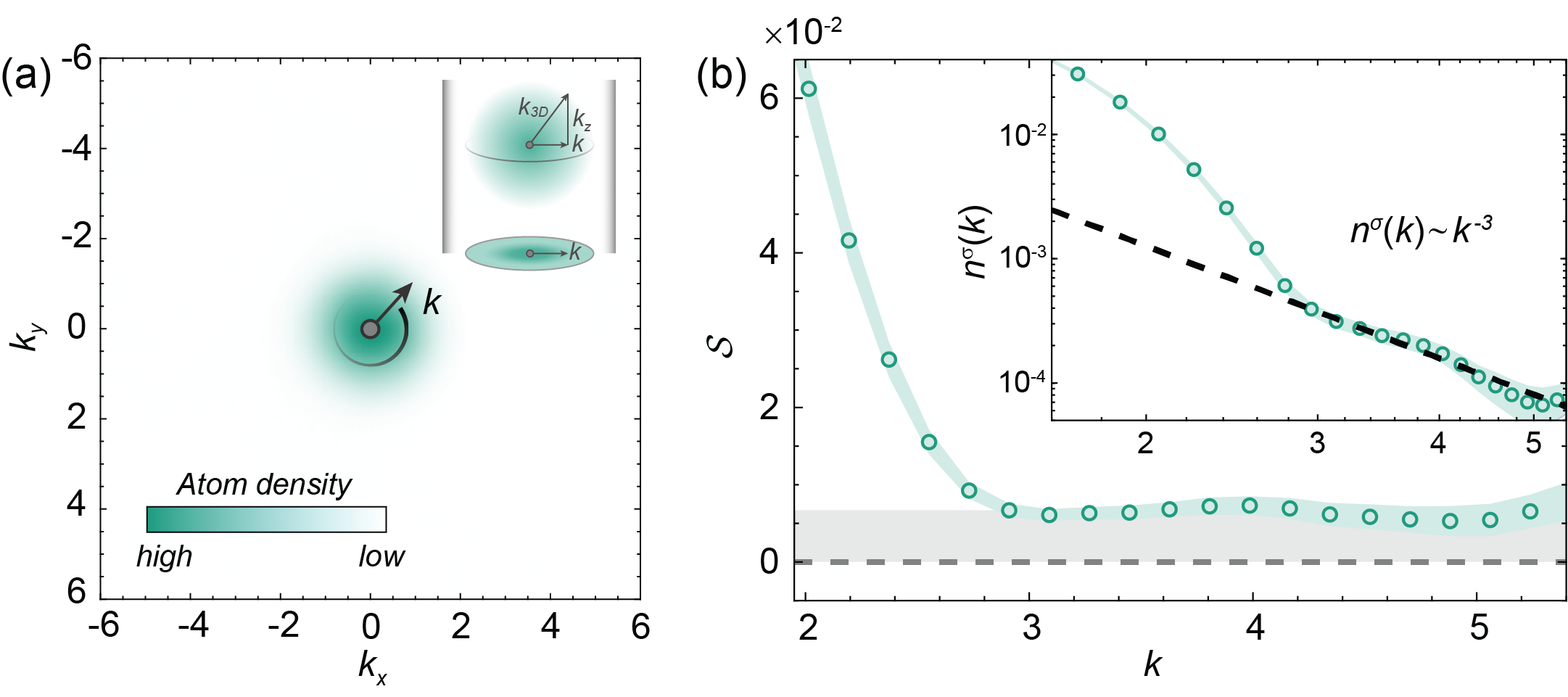}
	\centering
	\caption{\textbf{Measurement of the contact parameter from the momentum distribution}. (a) Momentum distribution of $^{173}$Yb atoms consisting of $N=6$ nuclear spin states after 4~ms time-of-flight expansion. {\color{blue}Dimensionless momentum $k$ here is normalized by the Fermi wave number $k_F$.} Note that momentum profile outside $k\simeq2.5$ has already been subtracted by the momentum profile of the spin-polarised gas ($N=1$) with the same total atom number. (b)  $\mathcal{S} = 2/\pi \cdot k^3n^\sigma(k)$ is plotted as a function of momentum $k$ in units of $k_F$. Inset is the momentum tail of azimuthally averaged atomic distribution $n^\sigma(k)$ in logarithmic scale with a dashed guideline of $n^\sigma(k)\propto k^{-3}$. The $n^\sigma(k)$ is normalized as $\int n^\sigma(k) 2\pi k dk=1$. The finite value of the contact is determined from the plateau of the $\mathcal{S}$ profile in the range of $k = 3$ to $4$. The green and grey shaded areas indicate the standard error and the measured value of contact, respectively.}	\label{fig2_method}
\end{figure*}

%In Fig\ref{fig1_method}, our schematic protocol for the measurement of pairwise contact is shown through the atomic cloud after the time-of-flight expansion. For an atomic cloud distributed isotropically in 3D, the weight of high-momentum tail proportional to $k_{3D}^{-4}$ can be obtained from $k^{-3}$-momentum tail of the absorption image recorded in the 2D plane, as rigorously shown in the method section. The key advantage of our protocol is that no transform gets involved resulting high signal-to-noise ratio. To obtain high signal-noise-ratio atomic profile, we average $200\sim600$ images as shown in Fig.1(a), and obtain the atomic distribution after radial-average. In this case, we normalize the atomic density as $\int n_{2D}(k) 2\pi k dk= I_0= (2\pi)^2$. To be noted, in the balanced $N$ components gases, due to the SU($N$) symmetry, the normalized atomic distribution of all $N$ spin components and single spin component are the same, denoted as $n(k)$ here.
%%\section*{High-precision measurement of high-momentum distribution}
%\paragraph*{\bf High-precision contact measurement : two steps}

%%\paragraph*{\bf Two-body contacts of SU($N$) fermions} %with two-body contact} 
%%A multi-component Fermi gas of $^{173}$Yb atoms is prepared in a three-dimensional trap 

\section{Bosonization and scaling of contacts in SU($N$) fermions}
The observed scalings of contacts can be qualitatively understood as follows. As
depicted in Fig.~\ref{fig1_scheme}, in a balanced SU($N$) gas with $N_0$ atoms per spin state, a single atom  with spin-$\sigma$ interacts with $(N-1) N_{0}$ atoms in the other $(N-1)$ spin components with spin-$\sigma'$ ($\sigma'\neq \sigma$) through the $s$-wave scattering.  
%%in a weakly interacting regime (see Fig.~\ref{fig1_scheme}), %
%%where three-body correlations are negligible. 
%Therefore,  
When interactions are spin-independent, each pair of atoms contributes %%equally to 
  an equal amount, $c_{\text{pair}}$, to the large momentum tail, $n^{\sigma}_{3D}(\vec{k})= \mathcal{C}_0/k_{3D}^4$, where $\vec{k}=(k_x,k_y,k_z)$ is a 3D momentum vector and its norm $k_{3D}=|\vec{k}|$ is much greater than $k_F$ and other microscopic momentum scales. 
%%the $s$-wave contact. 
In the low fugacity regime where three-body correlations are negligible, $\mathcal{C}_0=c_{\text{pair}} (N-1) N_{0}^2$, i.e., scaled with ($N-1$) when the number of spin, $N$, is tuned.  %Beyond the two-body contact, three-body contact arising from Efimov effect also plays a crucial role in the Bose gas and has been studied both in theory and experiment \cite{Fletcher:2017cu, Braaten2011Universal}. However, the three-body correlation is negligible in our system due to the small fugacity. [[3-body correlations and 3-body contact are different concepts]]
Correspondingly, if we consider the total momentum distribution, $n_{3D}(\vec{k})=\sum_\sigma n^\sigma_{3D}(\vec{k})$, we could define the total contact, $\mathcal{C}_{\text{SU(N})}=N\mathcal{C}_0= c_{\text{pair}}N(N-1)N_0^2$.  Dividing $\mathcal{C}_{\text{SU(N})}$ by $N_t^2$,  where $N_t=NN_0$ is the total particle number, we obtain that $\mathcal{C}_{\text{SU(N})}/N_t^2 = c_{\text{pair}} (1-1/N)$.
%%we observe that it approaches a constant with a scaling of $1/N$.  As explained later, this scaling is a direct evidence of the bosonization of SU($N$) fermions in the large $N$ limit. }

In our experiment, $p$-wave scatterings are negligible, as the current temperature regime is smaller than the barrier of the $p$-wave interaction~\cite{Fukuhara2007Degenerate}.  We, therefore, treat SU(1) fermions as non-interacting systems.  This is precisely the origin of the $1/N$ factor in the scaling of $\mathcal{C}_{\text{SU(N})}/N_t^2$ with $N$. The Pauli exclusion principle suppresses the $s$-wave scattering between two atoms with the same spin, as well as their contributions to the $s$-wave contact. To make a comparison, we consider spinless bosons with the same $N_t$, $T$ and the same scattering length, $a_s$. Though $c_{\text{pair}}$ is independent on statistics,  all $N_t(N_t-1)/2$ pairs of particles in spinless bosons contribute to contacts such that the high momentum tail is written as $n_{\text{B}}(\vec{k})= \mathcal{C}_{\text{B}}/k_{3D}^4$, where $\mathcal{C}_{\text{B}}=c_{\text{pair}} N_t(N_t-1)\approx c_{\text{pair}} N_t^2$ for large $N_0$, as the momentum distribution of identical particles doubles that of distinguishable particles. We obtain $\mathcal{C}_{\text{SU(N})}/N_t^2=\mathcal{C}_{\text{B}}/N_t^2(1-1/N)$, which shows that the $s$-wave contact of SU($N$) fermions approaches that of bosons with a $1/N$ scaling. Since $\mathcal{C}_0/N=(\mathcal{C}_{\text{SU(N})}/N_t^2)N_0^2$, we use the contact per spin, $\mathcal{C}_0/N$, to capture this scaling with a fixed $N_0$.

%temperature
 %The experimental verification of the contact scaling, however, still remains unexplored because the measurement of the contact in a weakly interacting regime requires high SNR in the momentum-state imaging. 

%\paragraph*{\bf Experimental procedure}
\section{Results}
The experiment starts with degenerate fermions prepared in a crossed hybrid optical dipole trap (ODT) consisting of far-detuned 1064~nm and 532~nm laser light. A six-component Fermi gas of $^{173}$Yb atoms, loaded from an inter-combination magneto-optical trap, is evaporatively cooled down to the temperature $\sim$100 nK in the ODT in 6~s. Along with the evaporation, an arbitrary spin mixture with $N$=1,2,...,6 is prepared using optical pumping and blasting processes~\cite{Song:2016ep}. Next, we exponentially ramp up ODT to the final trap depth in 60~ms resulting in sufficiently large trap frequencies (see Appendix for details). Finally, the momentum distribution after a 4~ms time-of-flight expansion is recorded in the $k_x$-$k_y$ plane by absorption imaging along the $z$ direction using the resonant imaging light of $^{1}$S$_{0}$$\rightarrow$$^{1}$P$_{1}$ transition. {\color{blue}We note that the measurement of contact at high momentum is negligibly affected by the finite expansion time~\cite{TOFnote}.} In Fig.~\ref{fig2_method}(b), a typical high-momentum tail is observed in the $\mathcal{S}$ profile after the systematic noise is filtered out~\cite{fringe:2019}.

%\paragraph*{\bf Measurement of % the high-momentum tail}
Our schematic protocol, for the high-precision measurement of the contact, is based on the momentum distribution of the atomic cloud after the time-of-flight expansion as shown in Fig.~\ref{fig2_method}. Typically, to measure contacts from the momentum distribution, the atomic profile recorded in the 2D plane, which represents a column integrated momentum distribution, needs to be inverse-Abel transformed to a 3D momentum distribution. However, inverse-Abel transform often intensifies measurement noise and exacerbates SNR because it involves a derivative of the atomic distribution, which inevitably limits the capability to detect contacts in a weakly interacting SU($N$) Fermi gas. To overcome this limitation, we developed a protocol to extract contacts directly based on the weight of the high-momentum tail from a 2D time-of-flight image without using the inverse-Abel transform.

 \begin{figure}%[tbp]
	\includegraphics[width=0.9\linewidth]{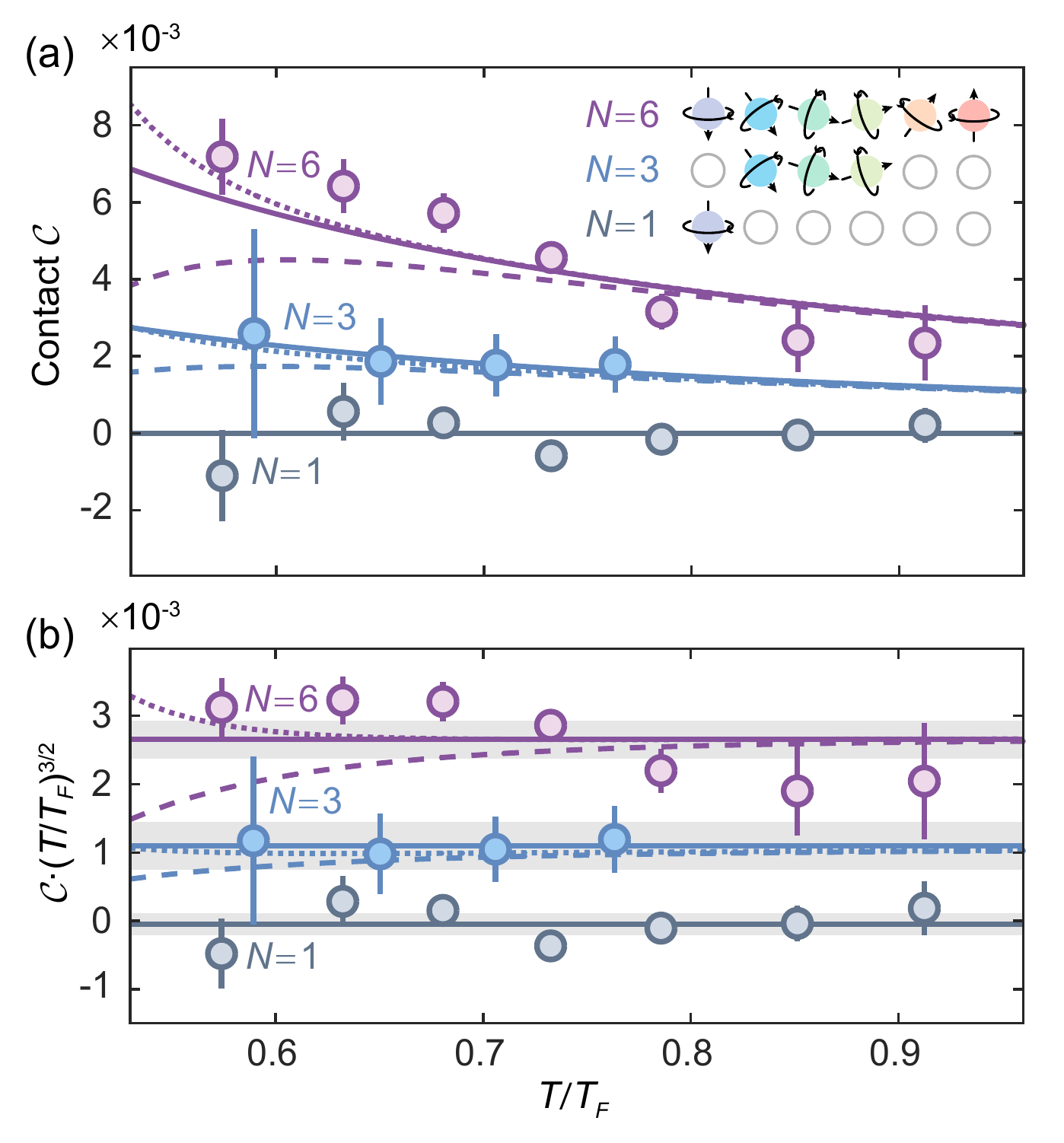}
	\centering
	\caption{\textbf{Temperature scaling of contacts in SU($N$) fermions}. (a) Contacts are measured at different temperatures in the SU($N$) Fermi gases. The error bars represent one standard deviation in the plateau area of $S$ profile. The solid curves are theoretical results multiplied by a factor of 7.5. Spin configuration of different SU($N$) gases are presented by arrows in spheres and the circle with empty inside indicates the absence of the spin state. Details of spin configurations and preparation of different SU($N$) gases are described in Appendix. %The result shows a good agreement with the temperature scaling 
	{\color{blue} Solid curves represent the results from the second order virial expansion, 
	$\mathcal{C}\propto (T/T_F)^{-3/2}$. Dashed and dotted curves represent results including corrections up to $b_3$ and $b_4$, respectively.} (b) Using the temperature scaling $\mathcal{C}\propto (T/T_F)^{-3/2}$, contacts of SU($N$) gases at different temperatures are collapsed to the Fermi temperature $T=T_F$. The solid line are means of collapsed contacts. The shaded grey area indicates the experimental uncertainty which consists of the standard error and the standard deviation of each point. {\color{blue} Dashed and dotted theoretical curves including high order corrections are no longer flat.} A horizontal error bar of $\pm$0.05$T/T_F$ is not shown for every data point.}
	\label{fig3}
\end{figure}

%For an atomic cloud distributed isotropically in 3D, the weight of the high-momentum tail proportional to $k_{3D}^{-4}$ can be obtained from $k^{-3}$-momentum tail of the absorption image recorded in the 2D $k_x$-$k_y$ plane, as rigorously proved in supplementary material. 
When $k_{3D}$ is much greater than the inverse of the harmonic oscillator length and other microscopic momentum scales, %%the momentum distribution 
$n^{\sigma}_{3D}(\vec{k})$ becomes isotropic in 3D and approaches $\mathcal{C}_0/k_{3D}^4 $.  Here, we have used $\mathcal{C}_0$ to distinguish the original definition of contact from the scaled one, $\mathcal{C}$, used in our experiment. To be noted, in a spin-balanced Fermi gas with $N$ components, atom density for each spin $n^{\sigma}_{3D}(k_{3D})$ is identical. Hereafter, it is normalized such that $\int n_{3D}^{\sigma}(k_{3D})d^3k_{3D} = 1$ in our experiment. Correspondingly, the column integrated momentum distribution, $n^{\sigma}(k) = \int_{-\infty}^{\infty} n^\sigma_{3D}(k_{3D}) dk_z$, which follows $\int n^\sigma(k) 2\pi k dk=1$. The momentum is normalized by the Fermi wave number $k_F = \sqrt{2E_Fm}/\hbar$ with the Fermi energy $E_F = \hbar \bar{\omega} (6N_0)^{1/3}$. Here $\bar{\omega}$ is the averaged trap frequency, $m$ is the mass of $^{173}$Yb and $\hbar$ is the reduced Planck constant. Contact $\mathcal{C}$ can be experimentally determined from the high momentum plateau of a term $\mathcal{S} = 2/\pi \cdot k^3n^\sigma(k)$ as follows (see Appendix),
%The key advantage of our protocol is that no transform gets involved resulting a high SNR ratio. The original Tan's contact $C_0$, associated with distribution at the high momentum regime, is defined as $\mathcal{C}_0 = \underset{k_{3D}\to \infty}{lim} k_{3D}^4 N(k_{3D})$ with a 3D wave number $k_{3D}$ and the atom number $N(k_{3D})$ at $k_{3D}$ in the momentum space. In our experiment, however, we measure the contact per particle $\mathcal{C}$ from the normalized integrated 2D column density $n(\vec{k})$ in the $k_x$-$k_y$ plane, where $\vec{k}$ is a two-dimensional vector normalized by the Fermi wave number $k_F = \sqrt{2E_Fm}/\hbar$ with the Fermi energy $E_F = \hbar \bar{\omega} (6N_0)^{1/3}$. Here $N_0$ is the atom number in one spin state, $\bar{\omega}$ the averaged trap frequency, $m$ the mass of $^{173}$Yb and $\hbar$  the reduced Planck constant. This contact $\mathcal{C}$ can be experimentally extracted from the high-momentum tail of $\mathcal{S} = 2/\pi \cdot k^3n(k)$ profile. Owing to the geometry of 3D and 2D wave number (see supplementary material for details), the contact defined here $\mathcal{C}$ has a simple relation with the original Tan's contact $\mathcal{C}_0$ as follows,
\begin{equation}
  \mathcal{C} = \lim_{k \to \infty} \mathcal{S}(k) = \lim_{k \to \infty}%
\frac{2}{\pi} \cdot k^3n^\sigma(k) =  \frac{\mathcal{C}_0}{(2\pi)^3 N_0 k_F}%
.
\end{equation}
Here, %contact 
$\mathcal{C}$ is naturally normalized by the atom number per spin $N_0$ and the Fermi wave number $k_F$. The key advantage of our protocol is that no transform gets involved resulting in a high SNR ratio. To further diminish the noise of the atomic profile, we typically repeat the measurement $\sim$100 times and obtain an averaged image as shown in Fig.~\ref{fig2_method}(a), and then azimuthally average the momentum distribution profile with $\pm$0.2$k_F$ moving average. To be noted, the value of the contact $\mathcal{C}$ measured by our protocol is in agreement with the result extracted from the 3D momentum distribution using the inverse Abel transform as described in Appendix.

%\paragraph*{\bf Removal of imaging noises} 
Because of the small scattering length of $^{173}$Yb, contacts in our SU($N$) gas are contained in the large momentum tail with an extremely small amplitude 
that is below a thousandth of the maximum cloud density. To extract such high-momentum tail from the subtle density profile, we first filter out the systematic noise (e.g. interference fringes, imaging light fluctuation) using  the statistical method. Our protocol is based on statistical image decomposition and projection methods using the data images as a basis set and compensating for unwanted fringes~\cite{fringe:2019}. Secondly, we compare the high-momentum tail of SU($N>$1) fermions with respect to non-interacting SU(1) gases, and extract the high-momentum tail of SU($N>$1) gases after subtracting the counterpart of SU(1). This allows us to systematically eliminate the diffraction effect arising from atoms. Note that for a SU(1) gas, we first separate the data set of SU(1) into two parts and analyse them using a similar procedure.

%\section*{Scalings of contacts with $T$ and $N$}

%\paragraph*{\bf Experimental results}
In Fig.~\ref{fig3}, we show the measured $\mathcal{C}$ at temperatures between $T/T_F=0.55$ and $T/T_F=1$ for SU($N$=1,3,6). We change the number of components, $N$=1,2,...,6, but keep the same number of atoms per spin component $N_0$=6.7$\times 10^3$ in a 3D harmonic trap with %%trapping 
frequencies  $(\omega_x,\omega_y,\omega_z)$$=2\pi\times$$(1400,750,250)$~Hz,  the averaged trap frequency $\bar{\omega} = (\omega_x\omega_y\omega_z)^{1/3}=$$2\pi\times$$640$~Hz and $k_F a_s\simeq$~0.3. We post-select data images according to the atom number and temperature with a tolerance of $\sim$0.1$T_F$. As expected, a spin-polarized SU(1) gas with negligible $p$-wave scatterings 
does not exhibit $k^4$ momentum tail within our experimental uncertainty while the finite contact is clearly observed for a SU(6) or SU(3) Fermi gas in Fig.~\ref{fig3}(a). Within the temperature regime we explored, 
 the contact increases as the temperature $T/T_F$ decreases.
 
\begin{figure}%[tbp]
	\includegraphics[width=0.9\linewidth]{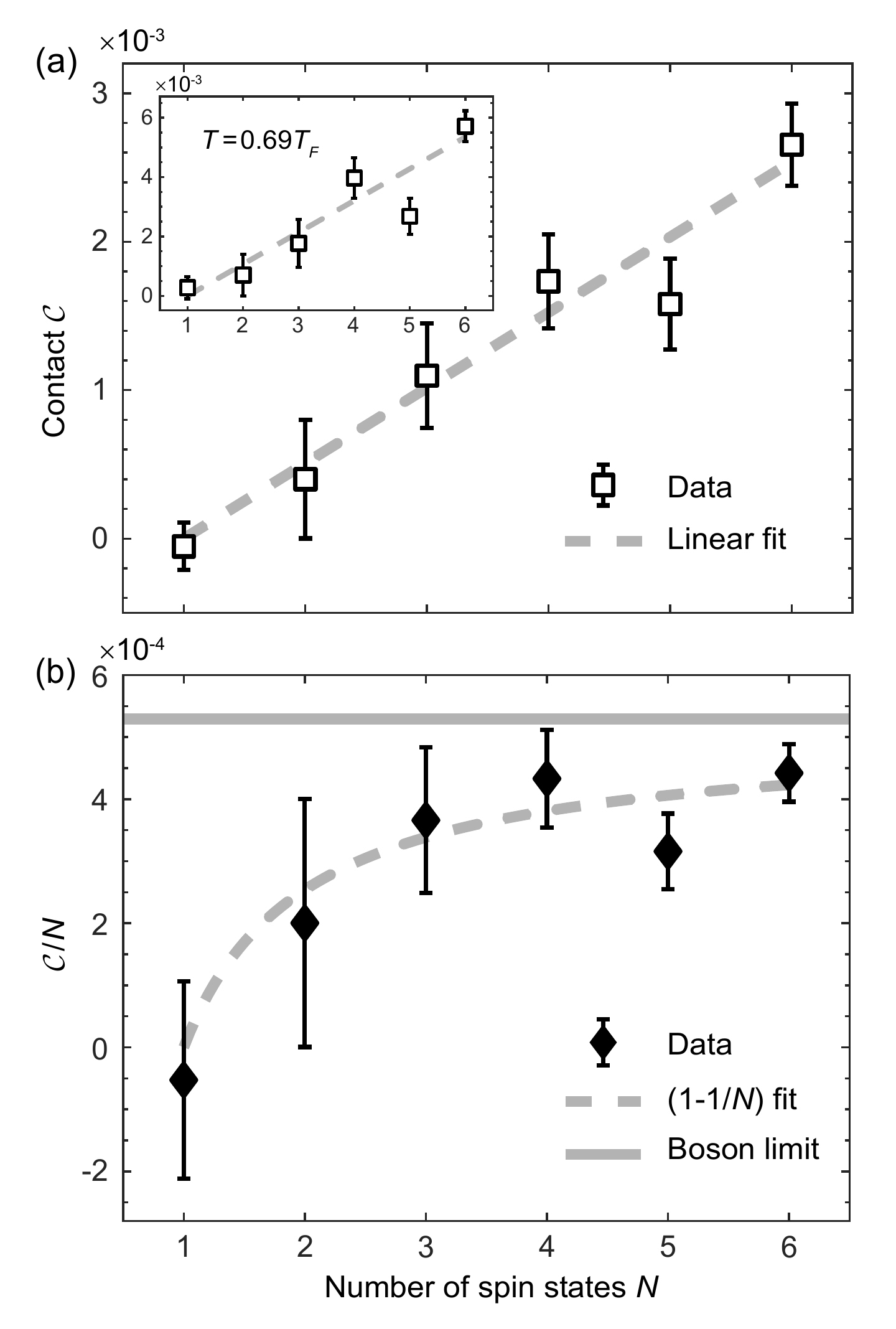}
	\centering
	\caption{\textbf{Bosonization in 3D SU($N$) fermions.} Scaling of the contact in SU($N$) fermions with tunable spin is shown. (a) Contacts scale linearly with the number of spin components in a SU($N$) Fermi gas. Mean values of collapsed contacts at $T=T_F$ of SU($N$) gases are plotted as a function of number of spin states. The dashed line is a linear fit to ($N-1$). Inset shows contacts at $T=0.69T_F$ . (b)  Contact per spin, $\mathcal{C}/N$, as a function of $N$. The dashed line indicates a (1-1/$N$) fit to the data.  The horizontal solid line denotes the theoretical result of trapped bosons, $ N_0\mathcal{C}_{\text{B}}/(8\pi^3k_FN_t^2)$, multiplied by 7.5.}
	\label{fig4}
\end{figure}

% \paragraph*{\bf Scaling of the contact with $N$ in SU($N$) fermions} 
In Fig.~\ref{fig4}, we test the scaling of the contact with the number of the spin components in SU($N$) Fermi gases.  We first collapse data points in Fig.~\ref{fig3}(a) to the Fermi temperature using $\mathcal{C}\propto(T/T_F)^{-3/2}$ shown in Fig.~\ref{fig3}(b). The results show that {\color{blue} the temperature dependence of $\mathcal{C}$ is qualitatively consistent with the $(T/T_F)^{-3/2}$ scaling, a prediction from the second order virial expansion (Appendix). However, high order virial expansions lead to corrections to the temperature dependence of the contact, and both the third and the fourth virial coefficients, $b_3$ and $b_4$, give rise to other powers of $T/T_F$ in the expression of the contact (Appendix). Such corrections are plotted in both Fig.~\ref{fig3}(a) and Fig.~\ref{fig3}(b). Since the result from the second order virial expansion qualitatively captures temperature dependance  and the current resolution in our experiment is not sufficient to accurately determine high order corrections, we empirically extract the mean values of collapsed contacts to the Fermi temperature, and further explore the dependence of the contact on $N$.} Fig.~\ref{fig4}(a) shows that $\mathcal{C}$ depends linearly on $(N-1)$, and  Fig.~\ref{fig4}(b) demonstrates that $\mathcal{C}/N\sim \mathcal{C}_0/N$ approaches a constant with a $1/N$ scaling. {\color{blue} Due to the smallness of $a_s/\lambda$ where $\lambda = \sqrt{2\pi\hbar^2/(mk_BT)}$ is the thermal wavelength, high order virial expansions do not change the $1/N$ scaling in the current parameter regime of our experiments (Appendix). } All results are consistent with the qualitative picture we previously provided.

%\paragraph*{\bf Theoretical results}

To measure $n^\sigma({k})$, we need to release atoms from the trap. Due to the absence of Feshbach resonance, interactions here cannot be turned off, unlike $^{40}$K for studying $s$-wave contacts of two-component fermions~\cite{Stewart:2010fy}. Interactions lead to complex expansion dynamics that are difficult to compute in theory. Therefore, it is illuminating to theoretically study contacts of trapped gases before the expansion. 
We compute contacts in the temperature regime explored in the experiments, $0.55 \le T/T_F\le 1.0$, where the second order virial expansion works well and high order virial expansion are negligible~\cite{Hu2011Universal,Sun:2015kf}. The virial expansion has been well established as a powerful tool to unfold fundamentally important principles using results at high temperatures,  for instance, the universality of fermions at resonance~\cite{ho2004high,Hu2009}. Here, the spirit is the same. The virial expansion does not only allow us to obtain quantitatively the $(1-1/N)$ scaling in three dimensions where exact solutions generically do not exist, but also reveal the crucial role of the large internal degree of freedom in bosonization.  %%We have verified that corrections from  in this temperature regime. 
%%We use the local density approximation to treat the inhomogeneity in the trap. 
We evaluate the local contact at the position $\vec{r}$ based on its local chemical potential $\mu_{loc}=\mu_0-V(\vec{r})$, where $\mu_0$ is the chemical potential at the center of the trap and $V(\vec{r})$ is the harmonic trapping potential. The total contact is obtained by integrating local contacts in the trap. {\color{blue} Using the second order virial expansion}, the contact  is written as 
$\mathcal{C}_0 = k_B T \frac{8\pi m}{\hbar^2}(\frac{k_BT}{\hbar\omega})^3z^2\frac{a_s^2}{2^{3/2}\lambda}(N-1)$.
$z = e^{\beta\mu_{loc}}$ is the fugacity. %%For the range of  temperature $T/T_F>0.55$, 
In this temperature regime, 
the chemical potential $\mu_0$ %%of the Fermi gas in a harmonic trap can be 
is well approximated by $\mu_0=-T/T_F\cdot \log(6(T/T_F)^3) E_F$~\cite{butts1997trapped}.  We obtain
%and find 
 %%Our theoretical results show the same temperature dependence, $\mathcal{C}\propto (T/T_F)^{-3/2}$, as the experimental data, though the absolute value of $\mathcal{C}$ is about $6.5$ times smaller in theory.  Moreover, the virial expansion leads to the same linear dependence of the contact per spin component on $N$ as found in experiments. These results, therefore, suggest that the scaling of contact with $T$ and $N$ remains unaffected by interaction effects in the expansion despite that the amplitude of contact itself has changed.  % for our experimental parameters %~\cite{Hu2011Universal,Sun:2015kf}. It is found that the scaled contact by the temperature collapses into the linear line as a function of $N$. Our measurement directly reveals the interaction effect in a multi-component Fermi gas with SU($N$) symmetry.
%%To better understand our measurement, we compute the two-body contact of the trapped gases before the expansion. 
%%We consider the thermodynamic potential in the grand-canonical ensemble of SU($N$) fermions. %%In the temperature regime explored by our experiment, $T/T_F>0.55$, the second order virial expansion works well for us to compute contacts. 
  \begin{equation}
	\mathcal{C} = \frac{\mathcal{C}_0}{(2\pi)^3 N_0k_F} %\propto 
	\approx
	\frac{ma_s^2}{6(2\pi)^{5/2}\hbar^2}\frac{E_F}{(T/T_F)^{3/2}}(N-1).
	\label{contactsunfermion}
\end{equation}
We observe that $\mathcal{C}$ scales with  $N-1$ % for the spin multiplicity $N$ and as $\mathcal{C}\propto(T/T_F)^{-3/2}$ for $T/T_F$ 
 and $(T/T_F)^{-3/2}$ in the high temperature regime. {\color{blue} In our experiments, $a_s/\lambda$ ranges between 0.06 and 0.08. In such weakly interacting regime, corrections from high order virial coefficients modify the temperature dependence but not the scaling with $N$ (Appendix). } Both scalings {\color{blue} with $T$ and $N$} are consistent with the aforementioned experimental results, suggesting that interactions during the expansion do not change the scalings of the contact with $T$ and $N$. 

%\begin{figure}[tbp]
%	\includegraphics[width=1\linewidth]{Fig5}
%	\caption{\textbf{Bosonization with increaseing $N$}.
%	   %We plot the total contact $\mathcal{C}_{SU(N)}$ normalized by the contact $\mathcal{C}_B$ of a bosonic gas with the same total atom number, showing bosonization in good agreement with the scaling $\mathcal{C}_{SU(N)}/\mathcal{C}_B\sim \mathcal{C}/N \sim (1-1/N)$.
%	    {\color{red} Contact per particle, $\mathcal{C}/N$, as a function of $N$. }
%	   A dashed line indicates a (1-1/$N$) fit to the data. {\color{blue} The horizontal line denotes the theoretical result of trapped bosons, $\mathcal{C}_{\text{B}}/N_t^2N_0^2$, multiplied by 6.5}. }
%	\label{fig5}
%\end{figure}

%\paragraph*{\bf Bosonization} We would like to emphasize that such linear scaling of the contact per spin component with $N$ reflects that the SU($N$) fermions bosonize with increasing $N$ in the low fugacity regime. This linear scaling tells us that the total contact is proportional to $N(N-1)N_0^2$, or equivalently, $(1-\frac{1}{N})N_t^2$. 
%If we consider a bosonic gas with the same total particle number, its total contact, $\mathcal{C}_{\text B}$, is proportional to $N_t(N_t-1)$, i.e, $\sim N_t^2$ in the large $N_t$ limit.  Thus, we conclude that 
Using the virial expansion, the $s$-wave contact of spinless boson is also obtained explicitly in the same low fugacity regime (see Appendix). {\color{blue} The second order virial expansion leads to } %We then find
\begin{equation} 
\mathcal{C}_{\text{SU(N)}}=N\mathcal{C}_0=\left(1-\frac{1}{N}\right) {\mathcal{C}_{\text B}}. \label{CBF}
\end{equation} 
%%This result can be rigorously proved by using the virial expansion to compute the contact of bosons in the same low fugacity regime (see Methods).
As contact is the central quantity to control the many-body system, %such $1/N$ deference between the contact of SU$(N)$ fermions and bosons 
 Eq.~(\ref{CBF}) is a direct proof of the bosonization  without resorting to any other quantities, such as
%In other words, bosonization can be %%directly demonstrated 
%unfolded from the high momentum tail without resorting to 
the full momentum distribution. {\color{blue} Corrections from the third and
the forth virial coefficients introduce an extra temperature-dependent factor in
Eq.~(\ref{CBF}) (Appendix).}

Whereas scalings of the measured contacts with $N$ and $T$ after the expansion are consistent with theoretical results of trapped gases, experimental results lie systematically above theoretical ones, the former about 7.5 times greater than the latter (see Appendix). It is interesting to note that such discrepancy was also observed in an experiment measuring contacts of a weakly interacting Bose-Einstein condensate of $^4$He atoms~\cite{Chang:2016fa}. Interactions remain finite during expansions in both cases.  It is, therefore, possible that interaction effects during the expansion lead to the aforementioned discrepancy. However, the current resolution limits our capability to measure the time dependence of contacts in the expansion, which by itself is an interesting question concerning the non-equilibrium dynamics of contacts. To avoid this issue and directly access contacts of trapped gases, an alternative scheme is the  Bragg spectroscopy without the expansion~\cite{Carcy:2019iw}.

\section{Discussions and conclusion}
%We have seen that the 

Whereas we have focused on {\color{blue} the high temperature regime,} %low fugacity regime, 
 it is will be interesting to explore the low temperature regime in the future. First, %with decreasing $T$, higher order virial expansions will become important. Though bosonization still exists,  scalings with $T$ and $N$ are expected to change.  
it is desirable to %study 
{\color{blue} experimentally resolve 
corrections to %how the difference between contacts of SU($N$) fermions and spinless bosons scales with some powers of 
the
$1/N$ scaling, %{\color{red} In particular, going beyond the first order virial expansion 
which 
will allow us to understand how three-body and four-body correlations may affect contacts and bosonization}. Second,
with further decreasing $T$ down to below the superfluid transition temperature, our scheme of measuring contacts without using the inverse-Abel transform will provide us with an even richer playground to study contacts of superfluids with the SU($N$) symmetry. It has been shown that contacts of superfluids are directly related to the superfluid order parameters \cite{Zhang2017, Yoshida2016}. It will be interesting to investigate how the interplay between the superfluid order and the SU($N$) symmetry may bring us new macroscopic quantum phenomena and new universal relations governed by contacts, which may reveal deep connections between short-range correlations and many-body coherence.

 It is worth mentioning that, due to the small scattering length of ytterbium, it is a difficult task to reach the exponentially small superfluid transition temperature. {\color{blue} Though using other species with larger scattering lengths could increase the superfluid transition temperature, there are currently two SU($N$) species with $N>2$ available in laboratories, $^{173}$Yb and $^{87}$Sr. Unfortunately,  the latter has an even smaller $a_s$ than $^{173}$Yb we are using now. As such, a more practical approach is to} implement advanced cooling schemes, such as those engineering thermal reservoirs to absorb entropy from the system of interest~\cite{Zhou2009, Zhou2011}. Such schemes have been recently used to access long-range antiferromagnetic order in optical lattices~\cite{Greiner2017}. {\color{blue} It has also been recently used to access an extremely low entropy per particle of bosons down to $0.002 k_b$ in optical lattices~\cite{Yuan2020}. Converting such entropy to the temperatures of ideal fermions leads to $T/E_F\sim 0.0004$, a temperature scale below the superfluid transition temperature of the weakly interacting  $^{173}$Yb ($T/E_F\sim 0.005$ for $N=2$). It is thus promising that superfluid transitions of SU($N$) fermions are accessible in the near future once the aforementioned cooling schemes or similar ones are implemented. In particular, the large spin degree of freedom may help SU($N$) fermions to achieve even lower temperatures than spinless bosons, due to the Pomeranchuk effect~\cite{Taie:2012tb}.  It is then expected} that utilizing similar schemes in SU($N$) fermions will lead physicists to a new playground for exploring bosonization and many other exciting phenomena in the presence of both a large internal degree of freedom and symmetry breaking.

{\color{blue} In addition to SU($N$) fermions, our high-sensitivity measurement of contact in a microscopic level is also useful for other systems. For instance, contacts has been theoretically established as a powerful tool to explore deep connections between short-range correlations and many-body physics in spin-orbit coupled systems~\cite{Peng2018,Zhang2018,Jie2018}. Since spin-orbit coupling is the fundamental mechanism behind topological quantum matters, we hope that our works will stimulate systematic studies of contacts in topological physics. }

\section*{Appendix}

\renewcommand{\thefigure}{A\arabic{figure}}
\setcounter{figure}{0} 

\begin{figure*}
	\includegraphics[width=0.94\linewidth]{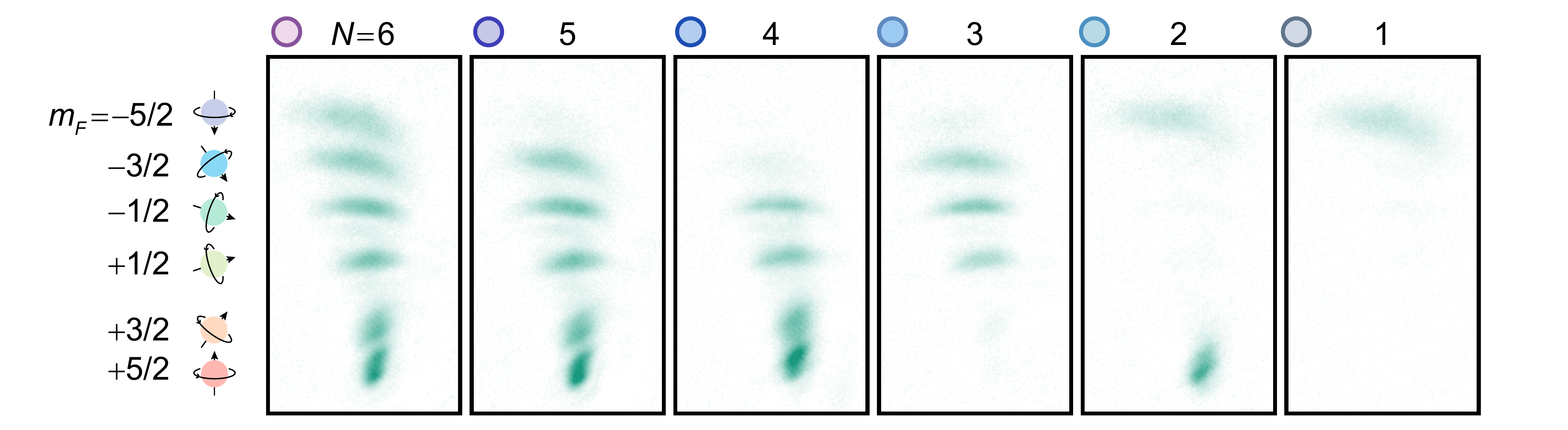}
	\centering
	\caption{\textbf{Fermi gases with tunable spin components.} Unwanted spin components are removed by short pulses of resonant $\sigma^+$ and $\sigma^-$ atomic transitions from $^{1}\text{S}_0$$\rightarrow$$^{3}\text{P}_1$ in an around 13.6~G magnetic field. From left to right, the number of spin states is prepared from $N=6$ to $N=1$. Optical Stern-Gerlach detection is used to monitor the spin configurations and split sub-clouds from top to bottom are $m_F=-5/2$ to $m_F=5/2$.}
	\label{figS1}
\end{figure*}

\paragraph*{\bf Preparation of SU($N$) gases}
 SU($N$) symmetric interaction in the ground state $^1$S$_0$ of $^{173}$Yb atoms emerges from the decoupling between nuclear spin and orbital angular momentum ($J=0$). Exploiting the energy splitting of the excited state in $^3$P$_1$ to our advantage, the narrow line-width transition $^1$S$_0(F=5/2)$$\rightarrow ^3$P$_1(F'=7/2)$, with wavelength $\lambda=556$~nm and natural line-width $\Gamma =  2\pi\times 181\ \text{kHz}$, is used as a blasting light to remove unwanted $m_F$ states of the ground manifold $^1$S$_0$.

The preparation starts with a gas of spin-balanced six $m_F$ states which is initially loaded in an optical dipole trap. 
A sequence of short pulses of $\sigma^\pm$ optical blasting light resonance to transition $m_F\rightarrow m_F\pm1$, is applied after the end of the evaporative cooling, where the temperature of atoms is $T$$\sim$$100\ \text{nK}$. The magnetic field of $13.6\ \text{G}$ is applied leading to a Zeeman splitting of $8.4\ \text{MHz}\sim46\ \Gamma$ between two adjacent $m_F$ states in the $^3$P$_1$ state. Take the preparation of a spin-balanced SU(2) gas as an example, we shine pulses of resonant blasting light with transitions $m_F=1/2 \rightarrow m_F'=3/2$, $m_F=3/2 \rightarrow m_F'=5/2$ with $\sigma^+$ polarization to remove positive $m_F=1/2$ and $m_F=3/2$ respectively, and $m_F=-1/2 \rightarrow m_F'=-3/2$, $m_F=-3/2 \rightarrow m_F'=-5/2$ with $\sigma^-$ polarization to remove negative $m_F=-1/2$ and $m_F=-3/2$ respectively, and the duration of each pulse is 5~ms. Following the similar method, arbitrary spin configuration of the SU($N$) ($N$=1,2...,6) gas can be prepared by the combination of $\sigma^+$ and $\sigma^-$ lights. The spin configurations of different SU($N$) gases used in the experiment are detected by optical Stern-Gerlach effect \cite{Song:2016ep} as shown in Fig.~\ref{figS1}.

\begin{figure}[b]
	\includegraphics[width=0.95\linewidth]{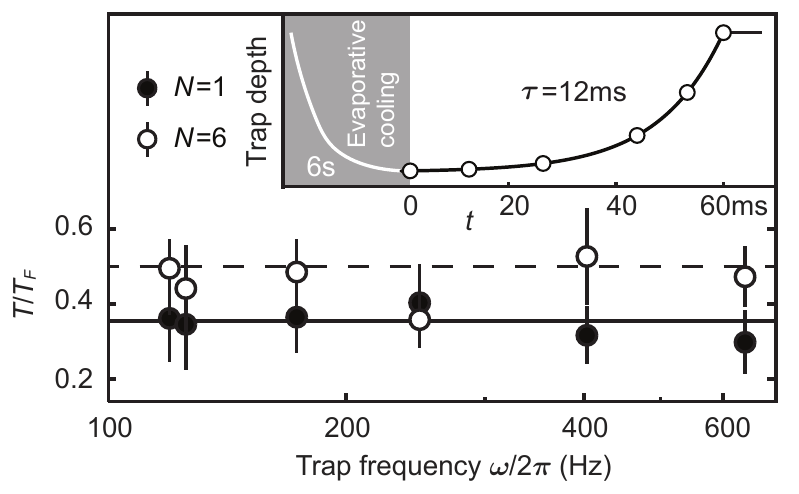}
	\centering
	\caption{\textbf{The ramp-up of the optical dipole trap.} $T/T_F$ of both non-interacting gases ($N=1$) and weakly interacting gases ($N=6$) are conserved during ramping up the optical dipole trap. The ODT is increased exponentially in $t_{ramp}=60$~ms with the time constant $\tau=12ms$ (inset). Gases with $N=1$ and $N=6$ components are initially prepared at $T=0.35T_F$ and $0.5T_F$, respectively.}
	\label{figS2_trap}
\end{figure}

Notably, we use optical pumping to prepare spin polarized gases ($N=1$) with different atom numbers. At the beginning of evaporation, we first optically pump most of atoms to the $m_F=5/2$ state using another optical pumping light 400~MHz red detuned from the resonance with $^1$S$_0$$\rightarrow$$^1$P$_1$ transition. The pumping pulse time is 300~ms. Note that we intentionally leave other spin states for the sake of the evaporative cooling. At the end of the evaporation, all the other remained spin states are removed by 556nm resonance light pulses similar to the procedure of SU($N$) gas preparation.
	
We further increase the trap depth to obtain large trap frequency, after the preparation of degenerate Fermi gases with different spin components at the temperature $\sim$100nK. $V(t)$, the trap depth of ODT is increased exponentially from the initial $V_i$ to the final trap depth $V_f$ in $t_{f}=60ms$ with a time constant $\tau=12ms$ as follows,
\begin{equation}
	V(t) = a e^{t/\tau} + b
\end{equation}
Where $a=(V_f-V_i)/(e^{t_f/\tau}-1)$ and $b=V_i-a$. We have experimentally tested that $T/T_F$ values of both non-interacting gases ($N=1$) and weakly interacting gases ($N=6$) are conserved during the ODT is ramped up as shown in Fig.~\ref{figS2_trap}.

\vspace{10pt}

\paragraph*{\bf Proof of the contact relation between $\mathcal{C}$ and $\mathcal{C}_0$} Different from the original approach using the inverse Abel transform to get 3D normalized distribution $n^\sigma_{3D}(k)$ from 2D TOF image \cite{Stewart:2010fy}, the method demonstrated here is more robust against noise because the contact is directly extracted based on the radial averaged atomic distribution $n^\sigma(k)$ from 2D TOF image, illustrated in Fig.2(a) in the main text. We calculate a term $\mathcal{S} = 2/\pi \cdot k^3n^\sigma(k)$ as a function of momentum $k$. The value of contact is experimentally extracted from the end tail of $\mathcal{S}$ profile. The contact $
\mathcal{C}$ is therefore determined as $\mathcal{C}=\underset{k\to \infty}{\text{lim}} \mathcal{S}$, which is slightly different from the original definition $\mathcal{C}_0$ \cite{Tan:2008ey, Tan:2008eg}. Contact defined here is naturally normalized by atom number per spin state $N_0$ and wave number $k_F$, and is associated with $\mathcal{C}_0$ as $\mathcal{C} = \mathcal{C}_0 / ((2\pi)^3N_0 k_F)$. Here is the detail of the proof. In a spherical symmetry system which is confirmed experimentally, we start the derivation from the original definition of the contact $\mathcal{C}_0$~\cite{Tan:2008ey, Tan:2008eg},\

%%{\color{blue} Internal note: I suggest that we use ${k}_{3D}$ , $\tilde {k}_{3D}$ for the momentum and the normalized momentum, respectively. The current notation may cause confusion about $k_F$: is it a normalized one? Also we should use  $\vec{k}_{3D}=(k_x, k_y, k_z)$, and $k_{3D}=|\vec{k}_{3D}|$, $\vec{k}= (k_x, k_y)$, $k=|\vec{k}|$ in the definition of $\mathcal{S}$ to avoid confusions. 

%%REPLY: so many tildes and variables make unreadable and complicated. Eventually I only use $k$, $k_{3D}$, $n$ and $n_{3D}$ and I normalize $k$ by $k_F$. Therefore $k$ and $\mathcal{C}$ are dimensionless while $\mathcal{C}_0$ has a unit of [1/L]. However to be noted, according to the original Tan's definition, the unit of $\mathcal{C}_0$ should be [1/L] while $\mathcal{I}= \Omega \mathcal{C}_0$ ($\Omega$ is the volume) has a unit of [1/L]. But many papers including Debby's and more recent ones from Martin and Christ already (mis-)use $\mathcal{C}_0$ instead of $\mathcal{I}$ as contact notation.	
%%}

\begin{figure}[t]
	\includegraphics[width=0.99\linewidth]{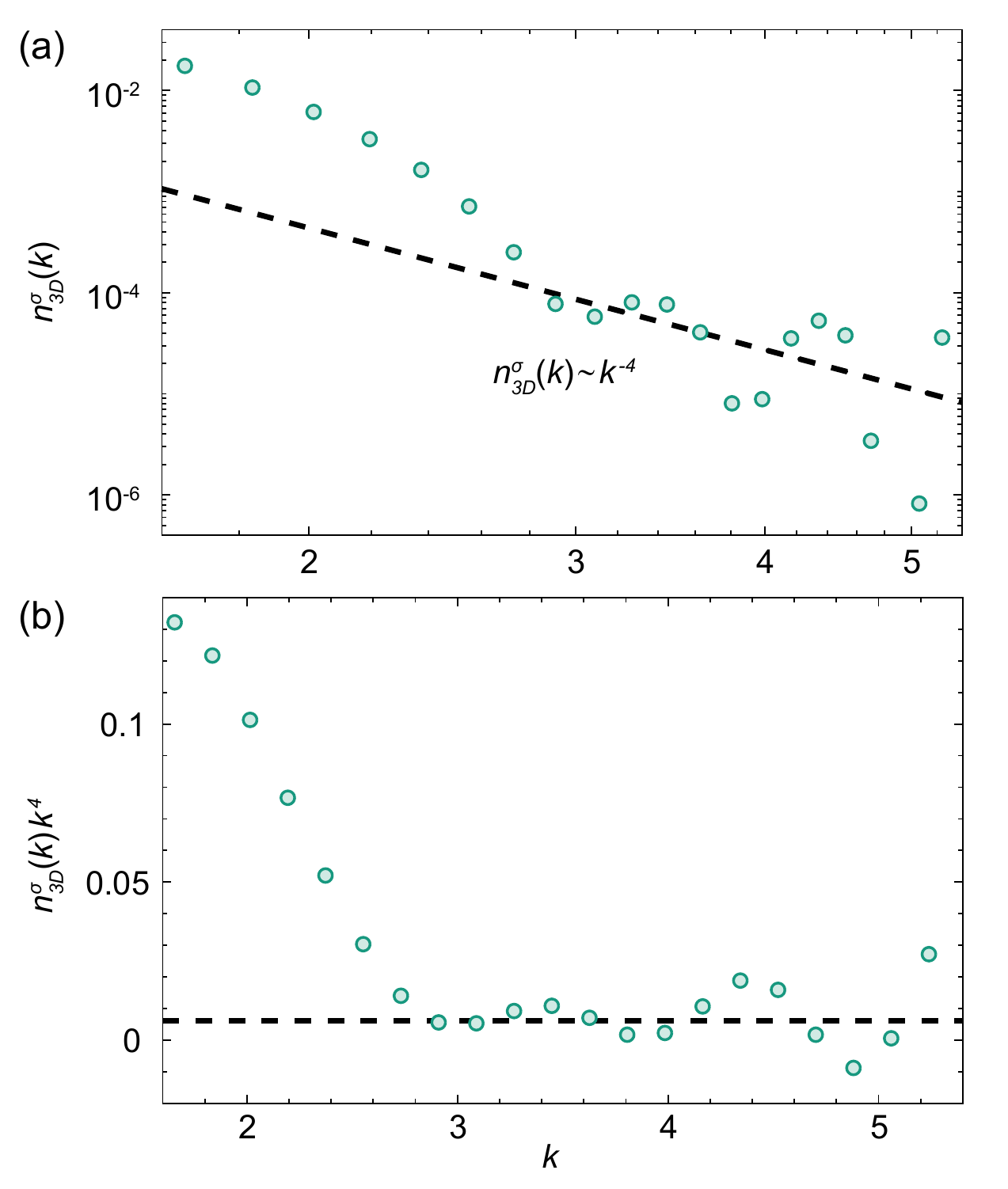}
	\centering
	\caption{\textbf{Extracting the contact from the 3D momentum distribution} (a) 3D momentum distribution of a SU(6) gas is obtained from the column integrated distribution (the same data in Fig.1 in the main text) using an inverse Abel transform. The 3D density in high momentum $k$ follows the power law $ n^{\sigma}_{3D}(k)\sim k^{-4}$, which gives the contact $\mathcal{C}=0.007(1)$ with $95\%$ confidence. (b) The contact is extracted from the tail of the term $n^{\sigma}_{3D}(k)k^4$ for $3<k<4$. The mean (one standard deviation in parentheses) is 0.006(3). The value of the contact extracted from the 3D momentum distribution is in agreement with the result of our method 0.0067(5) but has a larger uncertainty.}
	\label{figS3}
\end{figure}

\begin{equation}\label{eqn:contact1}
  \mathcal{C}_0 = (2\pi)^3 N_0k_F\lim_{k_{3D}\to \infty} k_{3D}^4 n^{\sigma}_{3D}(k_{3D}) 
\end{equation}
Where 3D wave vector $k_{3D} $ is normalized by $k_F$ and 3D density $n^\sigma_{3D}(k_{3D})$ is normalized such that $\int n^\sigma_{3D}(k_{3D})d^3k_{3D} = 1$.
The contact $\mathcal{C}$ defined in this article is written as,
\begin{align}\label{eqn:contact3}
\begin{split}
\mathcal{C}
& = \lim_{k \to \infty} \mathcal{S} \\
& = \lim_{k \to \infty} 2/\pi \cdot k^3n^\sigma(k) \\
& = 2/\pi \lim_{k \to \infty} k^3 \int_{-\infty}^{\infty} n^\sigma_{3D}(k_{3D}) dk_z\\
\end{split}
\end{align}
Here, we substitute the radial averaged atomic density with $n^\sigma(k) = \int_{-\infty}^{\infty} n^\sigma_{3D}(k_{3D}) dk_z$. From Eq.~(\ref{eqn:contact1}), for large $k_{3D}$, the 3D atomic density can be expressed as $n^\sigma_{3D} = \mathcal{C}_0/((2\pi)^3N_0k_F) \cdot k_{3D}^{-4} + \mathcal{O}(k_{3D}^{-5})$, in which $\mathcal{O}(k_{3D}^{-5})$ is the higher order term. Substitute $n^\sigma_{3D}$ into Eq.~(\ref{eqn:contact3}), with wave vector relation $k_{3D}^2  = k^2 + k_z^2$,

\begin{equation}\label{eqn:contact4}
\begin{split}
\mathcal{C}
& = \frac{2 \mathcal{C}_0}{\pi (2\pi)^3 N_0 k_F} \lim_{k \to \infty} k^3 \int_{-\infty}^{\infty} \frac{dk_z}{(k^2+k_z^2)^2} \\
& = \frac{\mathcal{C}_0}{(2\pi)^3N_0 k_F}  \\
\end{split}
\end{equation}

It is worth to note that we assume the momentum distribution is integrated over the  momentum $k_z$ in Eqs.~(\ref{eqn:contact3},\ref{eqn:contact4}). However, the true momentum distribution along the $k_z$ can be slightly perturbed  by atom-atom interactions during the expansion. If this is the case, the measured contact $\mathcal{C}$ may be proportional to the equation (6)  with an unknown factor.  To investigate this effect, we extract the value of contact from the three-dimensional density distribution using the inverse Abel transform in Fig.~\ref{figS3}, which does not require any approximation used in Eqs.~(\ref{eqn:contact3},\ref{eqn:contact4}). The contact measured from the three-dimensional density distribution is in good agreement with our result validating our detection method.

\paragraph*{\bf Theoretical model of contacts of SU($N$) fermions}
In the grand-canonical ensemble, the thermodynamic potential $\Omega$ for SU$(N)$ fermions can be expanded as a Taylor series at fugacity $z$ (virial expansion),
\begin{equation}
  \Omega=-k_B T Q_1(T) [N z + N b_{2}^{\text{ni}} z^2+ \frac{N (N-1)}{2}b_2(T,a_s) z^2+\dots],
  \label{<+label+>}
\end{equation}
where $Q_1(T)$ is the single particle partition function, $b_{2}^{\text{ni}}$ is the intraspecies second order virial coefficient which purely arises from particle statistics, and $b_2$ is the interspecies second order virial coefficient which typically depends on the scattering length and temperature. %[note that $b_2$ is temperature independent only for homogenenous systems.].
Using the adiabatic %sweep theorem 
relation~\cite{Tan:2008eg},

\begin{figure}[t]
	\includegraphics[width=0.99\linewidth]{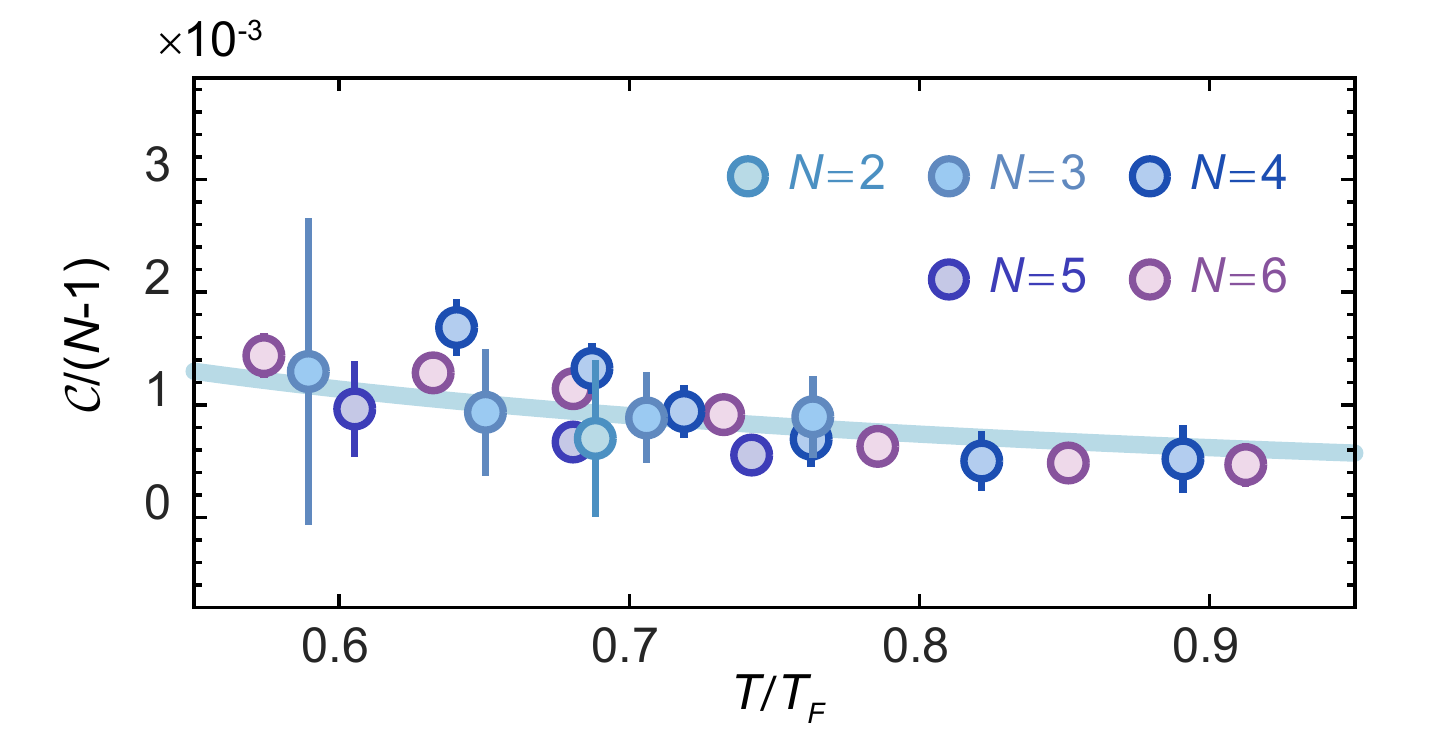}
	\centering
	\caption{\textbf{Scaled contact vs temperature} Contact of different SU($N$) gases are scaled on the $N=2$ components case by $\mathcal{C}/(N-1)$. The solid curve is the theoretical simulation multiplied by a factor of 7.5.}
	\label{figS4}
\end{figure}

\begin{equation}
  \left[\frac{\partial \Omega}{-a_s^{-1}}\right]_{T,\mu}=\frac{\hbar^2N}{8\pi m} \mathcal{C}_0,
  \label{<+label+>}
\end{equation}
we obtain an virial expansion of the contact from Ref.~\cite{liu2013virial},
\begin{equation}
  \mathcal{C}_0= k_B T \frac{4\pi m}{\hbar^2} Q_1(T)  z^2
  \frac{\partial b_2(T,a_s)}{\partial a_s^{-1}} (N-1).
  \label{contactfermion}
\end{equation}
Comparing the Taylor expansion of the grand canonical potential $\Omega$
  with the virial coefficient, we obtain
  \begin{equation}
    b_2(a_s,T)=Q_{\color{blue} 1,1}/Q_1,
    \label{<+label+>}
  \end{equation}
  where $Q_{\color{blue}1,1}$ is the partition function of two particles of different species
  in the anisotropic trap. 
  According to Ref.~\cite{gharashi2012three}, one could
  transform the problem, to a very good approximation at high temperature, 
  to a spherical harmonic trap with trapping frequency
  $\tilde{\omega}$ satisfies
  \begin{equation}
    3\tilde{\omega}^2=\omega_x^2+\omega_y^2+\omega_z^2.
    \label{<+label+>}
  \end{equation}
  Using the energy spectrum of two particles under isotropic harmonic confinement at any scattering length obtained from solutions in Ref.~\cite{Busch:1998vx} we can
  numerically determine the partition function as well as the derivative with respect to $a_s^{-1}$.
 
  According to local density approximation, {\color{blue} at high temperature
  where $k_B T \gg \hbar \bar{\omega}$},
  the virial coefficient for the trapped system can be related to that of the
  homogeneous system.
  \begin{equation}
  b_2\approx b_2^{\text{homo}}/2^{3/2},
    \label{}
  \end{equation}
  \begin{equation}
    Q_1\approx (k_B T/\hbar \bar\omega)^3,
    \label{}
  \end{equation}
  where $\bar\omega^3=\omega_x\omega_y\omega_z$.
  From Ref.~\cite{ho2004high}, one
  obtain
  $b_2^{\text{homo}}=-2a_s/\lambda$, where $\lambda$ is the thermal de Broglie length.
  Combining the equations above, under LDA, we obtain
\begin{equation}
  \mathcal{C}_0= k_B T \frac{4\pi m}{\hbar^2}  (k_B T/\hbar \bar\omega)^3
  z^2
  \frac{2a_s^2}{2^{3/2} \lambda } (N-1). 
  \label{2nd}
\end{equation}
In the high temperature limit, $z= {\color{blue} N_0(\hbar \bar\omega/ k_B T)^3}$.  The total contact of SU($N$) fermions is then written as 
\begin{equation}
 \mathcal{C}_{\text{SU(N)}}=N \mathcal{C}_0=
N^2_0 N(N-1) \frac{2\sqrt{2}\pi m }{\hbar^2} \frac{(\hbar \bar\omega)^3}{(k_B T)^2}\frac{a_s^2}{\lambda}.
 \end{equation}

\vspace{10pt}

\paragraph*{\bf Contact of single component Bose gas }
Applying the virial expansion to a single-component Bose gas,  the thermodynamic potential $\Omega_{\text B}$ at high temperatures is written as 
\begin{equation}
  \Omega_{\text B}=-k_B T Q_1(T) [ z_{\color{blue}\text B} + b_{2}^{\text{niB}}
  z_{\color{blue} \text B}^2+ b_2(T,a_s) z_{\color{blue}\text B}^2+\dots].
  \label{<+label+>}
\end{equation}
Here $b_2$ is the second order virial coefficient for two distinguishable particles, i.e., the same as that for the intraspecies $b_2$ for SU($N$) fermions, and 
$b_{2}^{\text{niB}}$ is a term that accounts for bosonic statistics which is independent of the scattering length. 
Using the adiabatic relation \cite{Tan:2008eg},
\begin{equation}
  \left[\frac{\partial \Omega_B}{-a_s^{-1}}\right]_{T,\mu}=\frac{\hbar^2}{8\pi m} \mathcal{C}_{\text B},
  \label{<+label+>}
\end{equation}
we obtain an virial expansion of the contact,
\begin{equation}
  \mathcal{C}_{\text B}= k_B T \frac{8\pi m}{\hbar^2} Q_1(T)  z^2
  \frac{\partial b_2(T,a_s)}{\partial a_s^{-1}}.
  \label{bosoncontact}
\end{equation}
Using ${\color{blue}z_{\text B}}= N N_0(\hbar \bar\omega/ k_B T)^3$, we obtain
\begin{equation}
  \mathcal{C}_{\text B}=(N_0 N)^2 \frac{2\sqrt{2}\pi m }{\hbar^2} \frac{(\hbar \bar\omega)^3}{(k_B T)^2}\frac{a_s^2}{\lambda}.
  \label{2ndb}
\end{equation}
%For comparison, from Eq.~\ref{contactsunfermion}, the total contact $N C_0$ for the SU(N) fermions is 
Compare $\mathcal{C}_{\text B}$ and $\mathcal{C}_{\text{SU(N)}}$, we obtain
\begin{equation}
 \mathcal{C}_{\text{SU(N)}}=\frac{N-1}{N} \mathcal{C}_{\text B}.
  \label{<+label+>}
\end{equation}
In the limit $N\to \infty$,  $\mathcal{C}_{\text{SU(N)}}$ approaches $\mathcal{C}_{\text B}$ with a scaling of $1/N$.

{\color{blue} \paragraph*{\bf High order virial expansions }
With decreasing temperature, high order virial expansions are required. We
consider corrections up to $b_4$, i.e., 
\begin{align}
  \Omega= &-k_B T Q_1(T) [N z + N b_{2}^{\text{ni}} z^2+ \frac{N (N-1)}{2}b_2(T,a_s) z^2] \nonumber\\ 
  &-k_B T(\Omega_3 z^3+\Omega_4 z^4\cdots),
\end{align}
where
{\color{blue}
\begin{align}
 % \Omega_2&=\binom{N}{2}Q_{1,1} +N Q_{2}-\frac{N^2}{2} Q_1^2\\
% &=\binom{N}{2}(Q_{1,1} -Q_1^2)+\frac{N}{2}(2Q_{2}-Q_1^2)
 % \\
  \Omega_3=& \binom{N}{3} Q_{1,1,1}+ 2\binom{N}{2}Q_{1,2}+N Q_{3}
  \nonumber\\&-  N Q_1 \Big(N
      Q_2+ \binom{N}{2}
  Q_{1,1}\Big)+N^3 Q_1^3/3  \\
  \Omega_4&= N(N-1) Q_{1,3}+\binom{N}{2}Q_{2,2} +\binom{N-1}{2} N Q_{1,1,2\nonumber}\\
&+\binom{N}{4}Q_{1,1,1,1} +NQ_4  \\
&- N Q_1 \Big(\binom{N}{3}Q_{1,1,1}+N Q_3+N(N-1)Q_{1,2}\Big)\nonumber\\
\nonumber
&-\frac{1}{2}\Big(\binom{N}{2}Q_{1,1}+N Q_{2}-N^2Q_1^2\Big)^2+\frac{1}{2}N^4Q_1^4;
\end{align}
%Here, $b_2^{\text{ni}}\equiv(2Q_{2}-Q_1^2)/Q_1$.
%$b_2^{\text{int}}\equiv(Q_{1,1}-Q_1^2)/Q_1$. 
  %The non-interacting contribution to $b_2$ is proportional to $N$, while
  %The interacting contribution to $b_2$ is proportional to $N(N-1)$.
Here, {\color{blue} $\binom{N}{k}=N!/(k!(N-k)!)$ is the binomial coefficient,} $Q_{n_1,n_2\dots,n_N}$ means the partition function of the
$N$-component fermionic system with $n_i$ number of fermions in the
$i$th component and the order of $n_1,n_2\dots,n_N$ does not matter because all
spins are equivalent. {\color{blue} For instance,} $Q_{1,2}$ means the partition function of {\color{blue} three 
%two-component Fermi gas 
fermions with one particle  in one spin state and two particles in another one. When each of the three fermions occupies a different  spin state, its partition function is denoted by $Q_{1,1,1}$. }%spin up particle and two spin down particles.
}

In our experiment, $a_s/\lambda$ is a small number between 0.06 and 0.08. In such 
a weakly interaction regime, where $(a_s/\lambda)^2\ll 1$, 
  $Q_{1,1,1}=Q_1^3+3(Q_{1,1}-Q_1^2)Q_1 +O\left( ( a_s/\lambda\right)^2)$. This
  result can be understood from the fact that the partition function of three
  fermions, all of which occupy different spin states, reduces to $Q_1^3$ in the non-interacting limit. Turning on a weak interaction, the leading correction is $3(Q_{1,1}-Q_1^2)Q_1\sim a_s/\lambda$, i.e, these three fermions can be decomposed to a pair of particles, which are interacting with the two-body interaction,  and a remaining one that does not interact with the pair. The factor of 3 comes from the three ways of such decomposition. We thus obtain
 % Here, the factor of 3 in the second term originates from 3 pairs of interaction.
  \begin{align}
  \Omega_3&=2\binom{N}{2}(Q_{1,2}-Q_1 Q_2-Q_{1,1}Q_{1}+Q_{1}^3)\\\nonumber
  &+N\left(Q_3-Q_1Q_2+Q_1^3/3\right).
    \label{<+label+>}
  \end{align}
 Similar to what we have seen in the second order virial expansion, $\Omega_3$ can also be separated into two parts, one is the result of non-interacting systems, and other other is the correction from interactions. We thus rewrite the above equation as  
  \begin{equation}
  \Omega_3/Q_1=N(N-1)b_3^{\text{int}}+N b_3^{\text{ni}}, 
    \end{equation}
  where  $b_3^{\text{int}}\equiv(Q_{1,2}-Q_1 Q_2-Q_{1,1}Q_{1}+Q_{1}^3)/Q_1$ and 
  $b_3^{\text{ni}}\equiv\left(Q_3-Q_1Q_2+Q_1^3/3\right)/Q_1$.
  Again, the non-interacting part of $b_3$ is proportional to $N$, while
  the contribution from interactions is proportional to $N(N-1)$. Higher powers of $N$, such as $N(N-1)(N-2)$, which is at least of the order of $(a_s/\lambda)^2$,  are high order corrections and thus are negligible in the weakly interacting regime.
 % Here the terms that is proportional to $N(N-1)(N-2)$ is of order $O(a_s/\lambda)^2$ and is omitted.

  The same analyses can be applied to $b_4$. %We can always approximate
  In the weakly interacting regime, $a_s/\lambda\ll 1$, the partition 
  functions of more than 3 spin components can be expressed  in terms of those
  of two and one spin components. We thus obtain
  \begin{align}
  -\frac{\Omega}{N k_B T Q_1}&=z+\frac{1}{2}z^2(\left( N-1 \right)
  b_2^{\text{int}}+b_2^{\text{ni}})\\\nonumber
 &+ \frac{1}{3}z^3(3(N-1) b_3^{\text{int}}+b_3^{\text{ni}}) \\\nonumber
  &+\frac{1}{4}z^4(4(N-1) b_4^{\text{int}}+b_4^{\text{ni}}) +\dots
\end{align}
where, $b_n$ are expanded in terms of $a_s/\lambda$,
\begin{align}
  b_2^{\text{ni}}&=-1/8, \\
  b_2^{\text{int}} &=-\frac{a_s}{\sqrt{2}\lambda}  +O(\frac{a_s}{
  \lambda})^2,\\
  b_3^{\text{ni}}&=1/27, \\
  b_3^{\text{int}} &=\frac{a_s}{3\sqrt{6}\lambda}  +O(\frac{a_s}{
  \lambda})^2, \\
  b_4^{\text{ni}}&=-1/64, \\
  b_4^{\text{int}} &=-(\frac{1}{64}+\frac{1}{12\sqrt{3}})\frac{a_s}{\lambda}  +O(\frac{a_s}{
  \lambda})^2.
\end{align}
{\color{blue}
To derive these expressions, we have adopted the results of 
SU(2) Fermi gas, i.e., $Q_{1,2}$, $Q_{1,3}$, etc. ~\cite{marcelino14}.
}
As expected, the dimensionless $b_n$ must be power series of the dimensionless parameter $a_s/\lambda$. Using $\Omega$, we obtain both the total particle number and the contact, 
\begin{align}
  N N_0&=-\frac{\partial \Omega}{ \partial \mu}=-\frac{\partial \Omega}{ \partial
  z}\frac{\partial z}{ \partial
  \mu}=-\frac{\partial \Omega}{\partial z} \frac{z}{k_B T},
  \label{eqnumber}
  \\
  \frac{N \hbar^2}{8\pi m} \mathcal{C}_0&=-\frac{\partial \Omega}{ \partial
  a_s^{-1}},
  \label{eqcontact}
\end{align}
both of which can be expressed in power series of $z$. Eliminating $z$, we obtain the contact as a function of $N$, $N_0$, and $T$, 
  {\color{blue}
\begin{align}
 & \mathcal{C}^{(4)}_{0}= 
  \mathcal{C}^{(2)}_{0} f_F(\frac{T}{T_F}; k_Fa_s)
\label{nnlo}
  \end{align}
  \begin{align}
 & f_F(\frac{T}{T_F}; k_Fa_s)=\big[1- (\frac{1}{9\sqrt{3}}-\frac{1}{48})(T/T_F)^{-3} \\\nonumber 
&+\big(\frac{1}{108 \sqrt{6}}+\frac{1}{576
\sqrt{2}}-\frac{5}{31104}-\frac{1}{216 \sqrt{3}}\big)(T/T_F)^{-6}  \\ \nonumber
&+\frac{(N-1) k_F a_s}{12\sqrt{2\pi}} (T/T_F)^{-5/2} 
+2\big(\frac{(N-1) k_F a_s}{12\sqrt{2\pi}}\big)^2 (T/T_F)^{-5} \\ \nonumber
&-(N-1) \frac{k_F a_s}{2\sqrt{\pi}}(\frac{7}{108 \sqrt{6}}-\frac{1}{72 \sqrt{2}})(T/T_F)^{-11/2}
+\dots\Big]
\end{align}
which is accurate up to $(T/T_F)^{-6}$. This requires us to include
contributions up to $b_4$ in Eq.~(\ref{eqcontact}) and up to $b_3$ in
Eq.~(\ref{eqnumber}). 
Here, the Fermi temperature is a function of $N_0$, $T_F = \hbar \bar{\omega}
(6N_0)^{1/3}/k_B$, $a_s/\lambda$ is replaced by $\frac{k_F a_s}{2\sqrt{\pi}} \sqrt{T/T_F}$.
$\mathcal{C}^{(2)}_{0}$ is the
previously obtained contact in the second order virial expansion,
{\color{blue}as shown in Eq.~(\ref{2nd}).} Since
{\color{blue}$\mathcal{C}^{(2)}_{0}\sim (N-1)$}, the last three terms in the above equation
leads to higher powers of $N$ in the scalings of the contact with $N$. However,
all of them are much smaller than the other terms due to the smallness of
$k_Fa_s$ and $a_s/\lambda$. Using our experimental parameters, the total
correction from these $(N-1)$-dependent terms for $N=6$ is up to $13\%, 10\%,
5\%$ for $T/T_F=0.55,0.69,$ and $1$ respectively. Thus, {\color{blue}
  $\mathcal{C}= \frac{\mathcal{C}_0}{(2\pi)^3 N_0 k_F}$  scales with $(N-1)$ in
our experiments.}

The slope of the linear scaling of $\mathcal{C}$ with $N-1$ is temperature
dependent. Theoretical results plotted in Fig.~\ref{figsup5} show that the slope increases with decreasing $T$. This has also been observed in experiments. As shown in Fig.~\ref{fig3}(b), our experimental resolution is not able to distinguish the small curvatures, which come from the small high order corrections to the linear scaling of $\mathcal{C}$ with $N-1$. As such, we use a linear fitting to obtain the slope. Again, the theoretical results of the slope are about 6-9 times of the
experimentally observed ones due to the expansion dynamics, as explained in the main text. 

If we only keep contributions up to $b_3$ in Eq.~(\ref{eqcontact}) and $b_2$ in
Eq.~(\ref{eqnumber}), 
\begin{align}
  \mathcal{C}^{(3)}_{0}&=
  \mathcal{C}^{(2)}_{0}\big[1- (\frac{1}{9\sqrt{3}}-\frac{1}{48})(T/T_F)^{-3} \\\nonumber 
  &+\frac{(N-1) k_F a_s}{12\sqrt{2\pi}} (T/T_F)^{-5/2}+\dots\Big]
  \label{nlo}
\end{align}
  Both results have been presented in Fig.~\ref{fig3} of the main text. 

  \begin{figure}[htp] 
\includegraphics[width=0.9\linewidth]{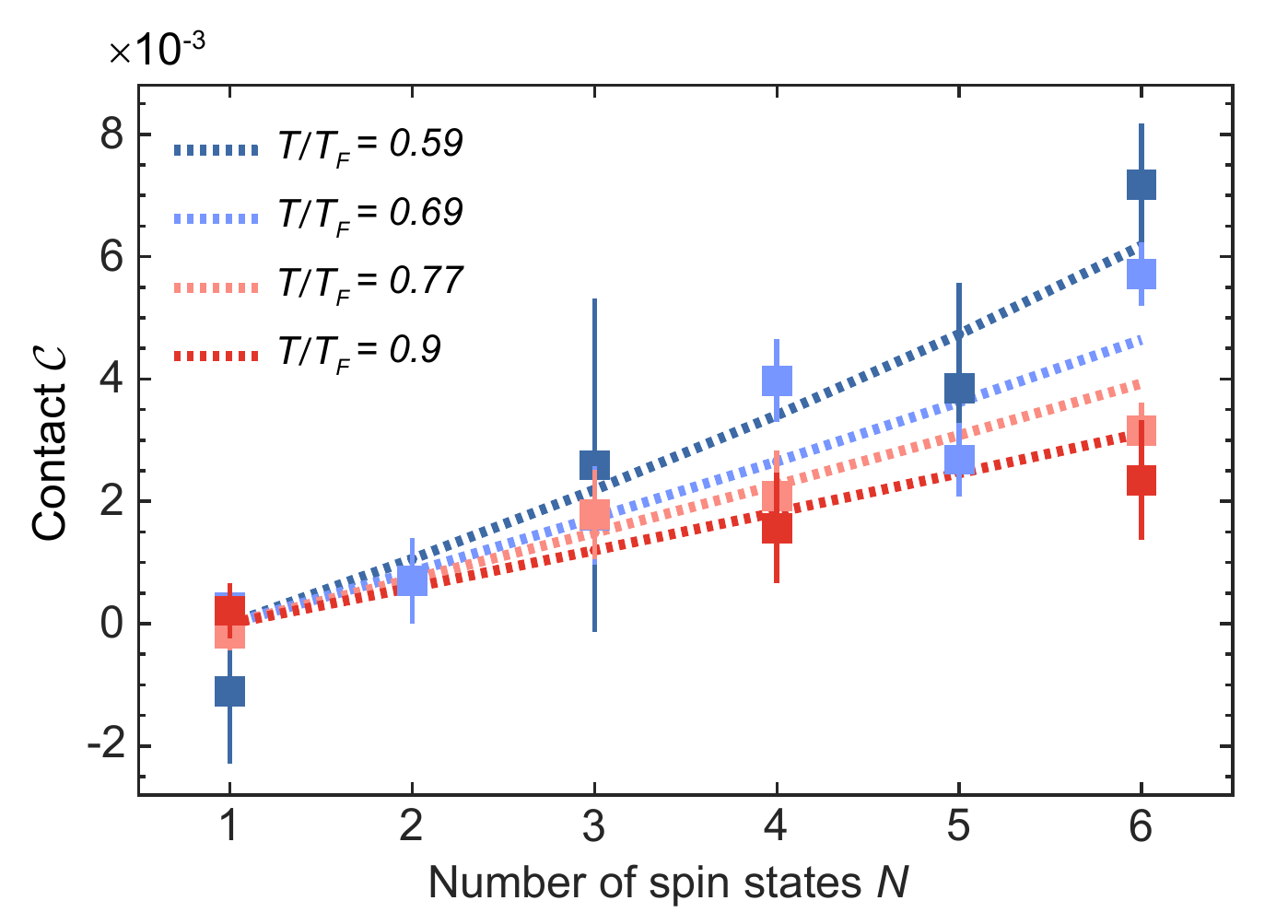}
\centering
\caption{ 
  \color{blue}
  Contact as a function of spin component $N$ at different temperatures $T/T_F$. Experimental results are denoted by boxes. Dotted lines represent theoretical results including corrections up to fourth order virial expansion. All theoretical curves have included a factor of 7.5.
}
	\label{figsup5}
\end{figure}
  
  \paragraph*{\bf{High order virial expansions for bosons}}
  The above analysis can be applied to bosons as well.
\begin{align}
 & \mathcal{C}^{(4)}_{\text B}=
  \mathcal{C}^{(2)}_{\text B}f_B(\frac{T}{T_F}; k_Fa_s),
  \end{align}
  \begin{align}
 & f_B(\frac{T}{T_F}; k_Fa_s)=\big[1+ (\frac{1}{9\sqrt{3}}-\frac{1}{48})N(T/T_F)^{-3} \\\nonumber 
    &+(\frac{1}{108
    \sqrt{6}}+\frac{1}{576 \sqrt{2}}-\frac{5}{31104}-\frac{1}{216 \sqrt{3}})N^2(T/T_F)^{-6}
    \\\nonumber
    &+\frac{k_F a_s}{6\sqrt{2\pi}}N^{5/6} (T/T_F)^{-5/2}+ \frac{(k_F
    a_s)^2}{36\pi}N^{5/3}(T/T_F)^{-5} \\ \nonumber
    &+(\frac{7}{54 \sqrt{6}}-\frac{1}{36 \sqrt{2}})\frac{k_F
    a_s}{2\sqrt{\pi}}N^{11/6}(T/T_F)^{-11/2}.
\dots\Big]
\end{align}
Here, {\color{blue}  $\mathcal{C}^{(2)}_{\text B}$ is the previously obtained
  contact of bosons in the second order virial expansion, as shown in Eq.~(\ref{2ndb}).
$T_F$ is defined to be the Fermi temperature of the corresponding Fermi
gas of a single component, $T_F = \hbar \bar{\omega} (6N_0)^{1/3}/k_B$, for convenience of comparison.
Using the results of both bosons and SU$(N)$ fermions, we obtain 
\begin{equation}
\mathcal{C}^{(4)}_{\text{SU(N)}}=N\mathcal{C}^{(4)}_{0}=\alpha(\frac{T}{T_F};k_Fa_s)(1-\frac{1}{N})\mathcal{C}^{(4)}_{\text B},
\end{equation}
where
$\alpha(\frac{T}{T_F};k_Fa_s)=\frac{f_B({T}/{T_F};k_Fa_s)}{f_F({T}/{T_F},k_Fa_s)}$.
The virial expansion works for SU$(N)$ fermions and bosons in the temperature
regimes $T   \gtrsim0.55 T_F$ and $T  \gtrsim 0.55 N^{1/3}T_F$,
respectively. At $T \gg 0.55 N^{1/3}T_F$, both $f_F(\frac{T}{T_F};k_Fa)$ and $f_B(\frac{T}{T_F};k_Fa)$ can be well approximated by 1. As such, $\mathcal{C}^{(4)}_{\text{SU(N)}}\rightarrow \mathcal{C}^{(4)}_{\text B}$  when $N\rightarrow \infty$. 
}
}

}
%\bibliography{scibib}

\begin{thebibliography}{32}
\bibitem{TOFnote}  The near-field effect arising from the initial size of the atomic cloud does not significantly affect the measurement of high-momentum tails. An atomic cloud with a radius $R$ after time-of-flight expansion has an near-field effect $dk\simeq Rm/(\hbar t_{tof})$ where $t_{tof}$ is the expansion time. In our experiment, the near-field correction, $dk/k\leq$~0.03, is negligible in the large-k regime ($k > 3 k_F$) where the contact is measured.
\end{thebibliography}

\bibliographystyle{Science}

\section*{Acknowledgments}
G.-B. J. acknowledges the generous support from the Hong Kong Research Grants Council and the Croucher Foundation through  GRF16311516, and GRF16305317, GRF16304918, GRF16306119, C6005-17G, N-HKUST601-17 and the Croucher Innovation grants respectively. Q. Z. is supported by NSF PHY 1806796.

%\section{}
%\subsection{}

\end{document}